\documentclass[%
reprint,
 showpacs,
 amsmath,amssymb,
]{revtex4-1}
\usepackage{graphicx}
\usepackage{dcolumn}
\usepackage{bm}

\usepackage{CJK}

\usepackage{color}
\usepackage{amssymb}
\usepackage{amsfonts}

\definecolor{Red}{rgb}{1,0,0}

\usepackage[colorlinks,urlcolor=blue,linkcolor=blue,citecolor=blue,anchorcolor=blue]{hyperref}
\makeatletter

\newcommand{\Rmnum}[1]{\expandafter\@slowromancap\romannumeral #1@}
\makeatother
\begin{document}

\preprint{APS/123-QED}

\title{Hierarchy of Genuine Tripartite Non-Gaussian Entanglement}


\author{Da Zhang$^1$}
\email{zhang1556433@sxnu.edu.cn}
\author{David Barral$^2$}
\email{david.barral@lkb.ens.fr}
\author{Yanpeng Zhang$^3$}
\author{Min Xiao$^4$}
\author{Kamel Bencheikh$^5$}
\affiliation{%
$^1$\mbox{School of Physics and Information Engineering, Shanxi Normal University, Taiyuan 030031, China} \\
$^2$\mbox{Laboratoire Kastler Brossel, Sorbonne Universit$\acute{e}$, CNRS, ENS-PSL Research University, Coll\`ege de France,}
\mbox{4 place Jussieu, F-75252 Paris, France} \\
$^3$\mbox{Key Laboratory for Physical Electronics and Devices of the Ministry of Education \& Shaanxi Key Lab of Information }
\mbox{Photonic Technique, School of Electronic and Information Engineering, Xi'an Jiaotong University, Xi'an 710049, China} \\
$^4$\mbox{Department of Physics, University of Arkansas, Fayetteville, Arkansas 72701, USA}
$^5$\mbox{Centre de Nanosciences et de Nanotechnologies, CNRS, Universit$\acute{e}$ Paris-Saclay, 91120 Palaiseau, France}\\
}%

\date{\today}

\begin{abstract}
\noindent
Triple-photon states generated by three-mode spontaneous parametric down-conversion are the paradigm of unconditional non-Gaussian states, essential assets for quantum advantage.
How to fully characterize their non-Gaussian entanglement remains however elusive.
We propose here a hierarchy of sufficient and necessary conditions for separability of the broad family of spontaneously-generated three-mode non-Gaussian states.
We further derive state-of-the-art conditions for genuine tripartite non-Gaussian entanglement, the strongest class of entanglement.
We apply our criteria to triple-photon states revealing that they are fully inseparable and genuinely entangled in moments of order 3n. Our results establish a systematic framework for characterizing the entanglement of triple-photon states and thus fostering their application in quantum information protocols.
\end{abstract}

\maketitle
Entanglement is a physical property describing the inseparability of quantum systems composed by multiple elements.
This core concept of contemporary physics dates back to 1935, when Einstein, Podolsky and Rosen proposed a \emph{gedanken} experiment criticizing the nonlocality of quantum mechanics and pointing out at a possible incompleteness of the theory \cite{einstein.pr.47.777.1935}.
Nowadays entanglement is prepared regularly in a number of physical systems \cite{Amico.rmp.80.517.2008}. For instance, entangled quadratures of electromagnetic fields are generated by single- and two-mode squeezing in parametric amplifiers and oscillators \cite{Heidmann.prl.59.2555.1987,zhangda.pra.96.043847.2017}.
These squeezed states are the cornerstone of multipartite entangled quantum networks and have greatly promoted the development of quantum optics \cite{yuan.pra.13.2226.1976,caves.pra.31.3093.1985,scully.quantumoptics.1999,agarwal2012quantum} and quantum information science \cite{gerd2007quantum,petz.quantuminformationprocessing.2007,braunstein2012quantum}.
They exhibit Gaussian statistics and their entanglement properties are completely specified by their covariance matrix \cite{vanloock.rmp.77.513.2005}.
However, it has been shown that non-Gaussian entangled states --inseparable states with non-Gaussian statistics-- are an essential resources for universal quantum computing \cite{l1oyd.prl.82.1784.1999,bartlett.prl.88.097904.2002,nielsen.prl.97.110501.2006,eisert.pra.82.042336.2010,eisert.pra.85.062318.2012}, demonstrating superior performance in many continuous variable protocols, such as quantum key distribution \cite{zubairy.pra.102.012608.2020}, quantum teleportation \cite{opatrn.pra.61.032302.2000,olivares.pra.67.032314.2003,dell.pra.76.022301.2007} and quantum metrology \cite{augusto.science.345.2014}.
Remarkably, the lack of passive separability of non-Gaussian entanglement has been recently pointed out as resource for quantum computational advantage \cite{Mattia2022resources}.
Non-Gaussian entangled states have been probabilistically created by photon addition and subtraction on Gaussian states along the last decades \cite{jeong.np.8.564.2014,morin.np.8.570.2014,ra.np.non.2020}, but a source of deterministic non-Gaussian entangled states was still missing.

Recently, three-mode unconditional non-Gaussian entangled states --triple-photon states (TPS)-- have been demonstrated in the microwave regime \cite{chang.prx.10.011011.2020}, and there is a great effort developing new platforms that produce them at optical wavelengths \cite{Moebius.oe.9.9932.2016,douady.23.2794.ol.2004,cavanna.pra.101.033840.2020,kangkang.aqt.35.2020}.
These new TPS are expected to extend the development of quantum optics and break the probabilistic nature induced by non-Gaussian operations in existing quantum information technologies.

Non-Gaussian entangled states are not only challenging to obtain in laboratories but are also hard to characterize.
Along the last two decades, efficient tools to detect multipartite Gaussian entanglement have been developed and experimentally tested countless times \cite{werner.prl.86.3658.2001,vanlook.pra.67.052315.2003}.
However, these criteria fail to detect fully-non-Gaussian entanglement \cite{kamel.prl.120.043601.2018,agust.prl.125.020502,zhang.pra.013704.2021}.
Several sufficient criteria have been proposed to detect the entanglement of known non-Gaussian states \cite{mm.prl.96.050503.2006,adam.pra.80.0523032009,hillery.pra.81.062322.2010}.
However, non-Gaussian states encompass a huge state space and necessary conditions of entanglement only arise when specific state features are taken into account \cite{Mattia.PRXQuantum.2.030204.2021}.
For TPS, sufficient conditions have been recently derived \cite{agust.prl.125.020502,tian.prapplied.18.024065.2022}.
These conditions are nevertheless not necessary --they do not fully or even unambiguously reveal the non-Gaussian entanglement of states--, only work in a limited parameter range, and are experimentally very demanding.
Thus, a systematic framework that fully characterizes TPS non-Gaussian entanglement and hence its operational usefulness \cite{Mattia2022resources}, sensitive at any parameter range and experimentally accessible is necessary.

In this Letter, we propose a hierarchy of full separability criteria for tripartite continuous variable states.
These criteria are a set of inequalities fulfilled by any three-mode biseparable state, based on linear combinations of experimentally-accessible high-order operators.
Violation of these inequalities at any hierarchy order is a sufficient condition for three-mode full inseparability.
Using high-order covariance matrices we find that these inequalities provide sufficient and necessary conditions for the full separability of spontaneously-generated three-mode non-Gaussian states, such as TPS.
This establishes a systematic framework for the characterization of TPS non-Gaussian entanglement.
Furthermore, based on our full separability conditions we derive a hierarchy of stringent conditions which rule out mixtures of biseparable states resulting in a criterion of genuine tripartite non-Gaussian entanglement.
Finally, we demonstrate the hierarchical entanglement structure of TPS in a experimentally relevant parameter space by means of numerical simulations.

We start our analysis by considering the interaction Hamiltonian describing the nondegenerate three-mode spontaneous parametric down-conversion
\begin{align}\label{eq1}
\hat{H}_I=i\hbar\kappa(\hat{a}_1^{\dag}\hat{a}_2^{\dag}\hat{a}_3^{\dag}\hat{a}_4-\hat{a}_1\hat{a}_2\hat{a}_3\hat{a}_4^{\dag}),
\end{align}
where $\kappa$ is the 3nd-order coupling constant.
The annihilation operators $\hat{a}_1$, $\hat{a}_2$, $\hat{a}_3$ and $\hat{a}_4$ describe respectively the three down-converted modes and the pump mode.
Under the evolution of vacuum or thermal states driven by the Hamiltonian (\ref{eq1}), it has been demonstrated theoretically \cite{agust.prl.125.020502,zhang.pra.013704.2021} and experimentally \cite{chang.prx.10.011011.2020} that third order is the lowest-order correlation of TPS.
Taking into account its symmetry and non-Gaussianity, TPS should present also $6th$- and $9th$-order quantum correlations, and even higher.
This means that the well-developed entanglement criteria involving second-order correlations are no longer applicable to TPS \cite{kamel.prl.120.043601.2018}.

To fully probe the high-order entanglement of the TPS, we introduce the high-order quadrature operators $\hat{q}^n_k=(\hat{a}_k^{\dag n}+\hat{a}^n_k)/2$ and $\hat{p}^n_k=i(\hat{a}_k^{\dag n}-\hat{a}^n_k)/2$ for the mode $k$, satisfying the commutation relation $[\hat{q}^n_k,\hat{p}^n_k]=i\hat{f}^n_k$, where the superscript $n$ is a positive integer representing the hierarchy index and the full expressions of $\hat{f}^n_k$ are given in the Supplementary Material \cite{triple.prl.2020}.
Likewise, the two-modes operators $\hat{q}^n_{lm}=(\hat{a}_l^{\dag n}\hat{a}_m^{\dag n}+\hat{a}^n_l\hat{a}^n_m)/2$ and $\hat{p}^n_{lm}=i(\hat{a}_l^{\dag n}\hat{a}_m^{\dag n}-\hat{a}^n_l\hat{a}^n_m)/2$ are defined for the modes $l$ and $m$, which follow the commutation relation $[\hat{q}^n_{lm},\hat{p}^n_{lm}]=i\hat{f}^n_{lm}$ \cite{triple.prl.2020}.
We also define the following linear combinations
\begin{align}\label{eq2}
 \hat{u}^n_{k}=g_{k,n}\hat{q}^n_k-\frac{1}{g_{k,n}}\hat{q}^n_{lm},~~~~~ \hat{v}^n_{k}=g_{k,n}\hat{p}^n_k+\frac{1}{g_{k,n}}\hat{p}^n_{lm},
\end{align}
for a given permutation \{$k$, $l$, $m$\} of \{1, 2, 3\}, where $g_{k,n}$ is an arbitrary real number.

The standard approach to witness tripartite entanglement is to examine the separability of the three possible bipartitions of the system.
Thus, let us consider the tripartite density operator $\rho = \sum_i \eta_i\rho^i_{k}\otimes\rho^i_{lm}$ with $\Sigma_i \eta_i=1$.
For brevity, we denote it as $\rho_{k,lm}$.
We derive in the Supplementary Material that for the biseparable state $\rho_{k,lm}$, the total variance of a pair of operators $\hat{u}^n_{k}$ and $\hat{v}^n_{k}$ satisfies the inequality \cite{triple.prl.2020}
\begin{align}\label{eq3}
\langle\Delta(\hat{u}^n_{k})^2\rangle+\langle\Delta(\hat{v}^n_{k})^2\rangle
\geq g_{k,n}^2f^n_k+\frac{f^n_{lm}}{g_{k,n}^2},
\end{align}
where $f^n_{k/lm}=\langle \hat{f}^n_{k/lm}\rangle$.
Thus we have the following theorem concerning full tripartite inseparability:

\emph{Theorem 1.--}Violation of the three inequalities
\begin{subequations}
\begin{align}
&F^n_1\equiv\langle\Delta(\hat{u}^n_{1})^2\rangle+\langle\Delta(\hat{v}^n_{1})^2\rangle-g_{1,n}^2 f^n_1 - \frac{f_{23}^n}{g_{1,n}^2}\geq0,\\
&F^n_2\equiv\langle\Delta(\hat{u}^n_{2})^2\rangle+\langle\Delta(\hat{v}^n_{2})^2\rangle- g_{2,n}^2 f^n_2 - \frac{f_{13}^n}{g_{2,n}^2}\geq0, \\
&F^n_3\equiv\langle\Delta(\hat{u}^n_{3})^2\rangle+\langle\Delta(\hat{v}^n_{3})^2\rangle-g_{3,n}^2 f^n_3 - \frac{f_{12}^n}{g_{3,n}^2}\geq0,
\end{align}\label{eq4}
\end{subequations}
with any hierarchy index $n$ is sufficient to confirm fully inseparable tripartite entanglement.

\emph{Proof.}--Inequality (\ref{eq4}a) is a necessary condition for the separability of the bipartition $1-23$.
Once inequality (\ref{eq4}a) is violated with any hierarchy index $n$, we can conclude that the state cannot be described by $\rho_{1,23}$. Similarly, the hierarchy of inequalities (\ref{eq4}b) and (\ref{eq4}c) are implied by biseparable states $\rho_{2,13}$ and $\rho_{3,12}$, respectively. Therefore, violating the three inequalities for any index $n$ negates all possible bipartitions, thus proving the full inseparability of the state.

The three hierarchical bounds given in Eq. (\ref{eq4}) must be fulfilled for the three biseparable states.
This naturally raises two questions: i) Do the states that violate these bounds necessarily possess non-Gaussian entanglement? ii) Are these bounds strong enough to ensure that the state satisfying the three hierarchical inequalities is fully separable?

The answer to the first question is negative.
Although the violation of inequalities (\ref{eq4}) depends on $\langle\hat{q}^n_k\hat{q}^n_{lm}\rangle$$>$0 and $\langle\hat{p}^n_k\hat{p}^n_{lm}\rangle$$<$0, these high-order moments do not identify the statistical properties of states.
In fact, we can easily find a three-mode Gaussian state that simultaneously violates the three inequalities in Eq. (\ref{eq4}), such as $\rho=1/2|\alpha\rangle\langle \alpha|\otimes|\psi(r)\rangle\langle\psi(r)|+1/2|\psi(r)\rangle\langle\psi(r)|\otimes|\alpha\rangle\langle \alpha|$, where $|\alpha\rangle$ is a coherent state with classical complex amplitude $\alpha$ and $|\psi(r)\rangle$ is a two-mode squeezed vacuum state with squeezing parameter $r$ \cite{triple.prl.2020}.
This suggests that these criteria involving high-order moments have downward compatibility \cite{shchukin.pra.93.032114.2016,Shchukin.pra.90.012334.2014}, i.e., they can diagnose some Gaussian entanglement.

So far, two types of continuous variable states have been identified regarding their statistical features: Gaussian and non-Gaussian states  \cite{jeong.np.8.564.2014,morin.np.8.570.2014,ra.np.non.2020,zhang.pra.013704.2021}.
We find that the inequalities (\ref{eq4}a)-(\ref{eq4}c) are indeed sufficient and necessary conditions for the full separability of the high-order moments of spontaneously-generated non-Gaussian states, such as TPS.
The proof steps are as follows.

We first collect the high-order quadrature and built-up operators in the vector $\hat{R}^n=(\hat{q}^n_k,\hat{p}^n_k,\hat{q}^n_{lm},\hat{p}^n_{lm})^T$ and write the commutation relations as
\begin{align}\label{eq5}
[\hat{R}^n_i,\hat{R}^n_j]=i\Omega^n_{ij}, ~~~i,j=1,\dots,4,
\end{align}
where $\Omega^n = i\hat{f}^n_ks_y\oplus i\hat{f}^n_{lm}s_y$ and $s_y$ represents the $y$ Pauli matrix.
Analogous to the Gaussian states case, the high-order covariance matrices $V^{n}$ is defined as $V^n_{ij}=\langle\Delta\hat{R}^n_i\Delta\hat{R}^n_j+\Delta\hat{R}^n_j\Delta\hat{R}^n_i\rangle/2$, where $\Delta\hat{R}^n=\hat{R}^n-\langle \hat{R}^n\rangle$ and the high-order local moment $\langle \hat{R}^n\rangle=\mathrm{tr}[\hat{R}^n\rho]$, with $\rho$ being the density operator of the system.
Then we have $V^n_{ij}+i\langle\Omega^n_{ij}\rangle/2
=\langle\hat{R}^n_i\hat{R}^n_j\rangle$, where the commutation relation (\ref{eq5}) and the property $\langle\hat{R}^n\rangle=0$ are used \cite{triple.prl.2020}.
Hence we have the following statement of the uncertainty principle
\begin{align}\label{eq6}
V^{n}+\frac{i}{2}\langle\Omega^n\rangle \geq 0.
\end{align}
Every physical state that satisfies $\langle\hat{R}^n\rangle=0$, i.e., with zero local moments,  must conform to this inequality.

$V^n$ is by definition a symmetric matrix, which can be divided into $2\times2$ sub-blocks
\begin{align}\label{eq7}
V^n=\left(
    \begin{array}{cc}
      A_k & C_{k-lm} \\
      C_{k-lm}^T & B_{lm} \\
    \end{array}
  \right),
\end{align}
where $A_k$ and $B_{lm}$ are local high-order covariance matrices related respectively to the subsystems $k$ and $lm$, and $C_{k-lm}$ represents their correlation.
Using Williamson's theorem and a suitable singular value decomposition \cite{horn2012matrix}, we can always transform Eq. (\ref{eq7}) into the following standard form \cite{triple.prl.2020}
\begin{align}\label{eq8}
V^n_1=\left(
  \begin{array}{cccc}
    n_1 & 0 & s_1 & 0 \\
    0 & n_2 & 0 & s_2 \\
    s_1 & 0 & m_1 & 0 \\
    0 & s_2 & 0 & m_2 \\
  \end{array}
\right),
\end{align}
where the matrix elements satisfy the relations
\begin{align}\label{eq9}
\mathfrak{n}_2\mathfrak{m}_1=\mathfrak{n}_1\mathfrak{m}_2,~~
2(|s_1|-|s_2|)=\sqrt{\mathfrak{n}_1\mathfrak{m}_1}-\sqrt{\mathfrak{n}_2\mathfrak{m}_2},
\end{align}
with $\mathfrak{n}_i=2n_i-f^n_k$ and $\mathfrak{m}_i=2m_i-f^n_{lm}$ ($i=1,2$).

Note that these high-order covariance matrices from $V_1^1$ to $V_1^n$ only describe the correlation information between subsystem $k$ and $lm$.
To fully characterize three-mode non-Gaussian state with zero local moments, such as TPS, we need three sets of hierarchical higher-order covariance matrices.
In addition, local symplectic transformations do not affect the separability of $V^n$, which implies that states with the same three sets of hierarchical standard forms (\ref{eq8}) have the same entanglement structure.
With these preliminaries, we now present the main theorem about the separability of $V_1^n$.

\emph{Theorem 2:} The necessary and sufficient condition for the separability of $V^n_1$ is that the hierarchy of operators
\begin{align}\label{eq10}
\hat{u}^n_{k}&=g_{k,n}\hat{q}^n_k-\frac{s_1\hat{q}^n_{lm}}{g_{k,n}|s_1|},~~\hat{v}^n_{k}=g_{k,n}\hat{p}_k+\frac{s_2\hat{p}^n_{lm}}{g_{k,n}|s_2|},
\end{align}
satisfy the inequality (\ref{eq3}), where $g_{k,n}^2=\sqrt{\mathfrak{m}_1/\mathfrak{n}_1}$.

\emph{Proof.}--Inequality (\ref{eq3}) is already a necessary condition for the separability of $k-lm$, so we only need to prove its sufficiency.
Substituting $\hat{u}^n_k$ and $\hat{v}^n_k$ of Eq. (\ref{eq10}) into inequality (\ref{eq3}) and using the standard form $V^n_1$, we obtain the inequality $g_{k,n}^2 (\mathfrak{n}_1+\mathfrak{n}_2)+g_{k,n}^{-2}(\mathfrak{m}_1+\mathfrak{m}_2)-4|s_1|-4|s_2|\geq0$. Combined with Eq. (\ref{eq9}), one finds
\begin{align}\label{eq11}
2|s_i| \leq \sqrt{\mathfrak{n}_i\mathfrak{m}_i}.
\end{align}
Since the uncertainty principle is invariant under local standard transformations, the standard form $V^n_1$ always satisfies Eq. (\ref{eq6}), which can be further reduced to $\det(V_1^n+i\langle\Omega^n\rangle/2)\equiv (f^n_k)^2(f^n_{lm})^2\det(V^n_G+i\langle\Omega\rangle/2)\geq0$, where $\Omega=\text{diag}(J^1_k,J^1_k)$ and $V^n_G$ is given in \cite{triple.prl.2020}.
This is equivalent to inserting a normalization coefficient into the commutation relations and it does not affect the separability of $V^n_1$.
The standard form $V^n_G$ and the uncertainty principle $V^n_G+i\langle\Omega\rangle/2\geq0$ suggest that $V^n_G$ can be regarded as a two-mode Gaussian state in the canonical quadratures.
Inequality (\ref{eq11}) ensures that every matrix $V^n_G-1/2$ is semi-positive definite, implying that all Gaussian states represented by $V^n_G$ are separable \cite{simon.prl.84.2726.2000,duan.prl.84.2722.2000}, which demonstrates that the original states $V^n_1$ are separable. This completes the proof of theorem 2.

Thus, we have the following result: \emph{A non-Gaussian state represented by $3n$ high-order covariance matrices $V^n_1$ is fully inseparable iff three pairs of hierarchical operators violate inequalities (\ref{eq4}a)-(\ref{eq4}c), respectively, with any hierarchy index $n$.}
Theorem 2 is a natural extension of the Duan criterion \cite{duan.prl.84.2722.2000} to the tripartite high-order correlation system.
Notably, this framework is inherently scalable and can be easily extended to $N$-partite systems.

Full inseparability is however not the more general form of multipartite entanglement \cite{eisert.njp.8.51.2006,shalm.np.9.1.2012}. It can
only exclude any biseparable case rather than the general one in which the state can be described as a mixture
\begin{align*}
\rho  &= P_1\sum_i \eta^{(1)}_i\rho^i_{1,23}+P_2\sum_t \eta^{(2)}_t\rho^t_{2,13}+P_3\sum_j \eta^{(3)}_j\rho^j_{3,12},
\end{align*}
where $\Sigma_iP_i=1$.
If a tripartite state can not be described by this equation, it is said to be genuinely entangled.
Genuine entanglement and full inseparability are equivalent for pure states, but for mixed states, the former is more strict than the latter
\cite{teh.pra.90.062337.2014,shalm.np.9.1.2012}.
Substituting the above equation into inequality (\ref{eq4}), we derive a hierarchy of genuine tripartite entanglement criterion in the Supplementary Material \cite{triple.prl.2020}, which can be stated as follows:



\emph{Theorem 3:} A tripartite state is genuinely entangled if the inequality
\begin{align}\label{eq13}
W_n &\equiv F^n_{1}+F^n_{2}+F^n_{3} +4\langle\hat{q}^n_1\hat{q}^n_{23}\rangle_\rho-4\langle\hat{p}^n_1\hat{p}^n_{23}\rangle_\rho \nonumber \\
& +2(\langle\hat{a}^{\dag n}_{1}\hat{a}^{ n}_{1}\rangle_\rho+\langle\hat{a}^{\dag n}_2\hat{a}^{n}_2\rangle_\rho\langle\hat{a}^{\dag n}_3\hat{a}^{n}_3\rangle_\rho) \geq 0
\end{align}
is violated for any hierarchy index $n$.


Let us now discuss the main features of the proposed criteria.
For any tripartite continuous variable state, violations of criteria (\ref{eq4}) and (\ref{eq13}) are sufficient to confirm fully inseparable and genuine tripartite entanglement, respectively.
When the hierarchy index $n=1$, the matrix elements in block $C_{k-lm}$ can be decomposed into the superposition of 3rd-order standard moments --coskewness--, i.e., $\langle\hat{p}^1_k\hat{p}^1_{lm}\rangle=\langle\hat{p}^1_k\hat{p}^1_{l}\hat{q}^1_{m}\rangle+\langle\hat{p}^1_k\hat{q}^1_{l}\hat{p}^1_{m}\rangle$ and $\langle\hat{q}^j_k\hat{q}^j_{lm}\rangle=\langle\hat{q}^1_k\hat{q}^1_{l}\hat{q}^1_{m}\rangle-\langle\hat{q}^1_k\hat{p}^1_{l}\hat{p}^1_{m}\rangle$, recently measured experimentally for TPS \cite{chang.prx.10.011011.2020}.
Inequality (\ref{eq11}) indicates that even if there are non-Gaussian correlations among the three modes, they may be separable, which implies that non-zero high-order standard moments are a necessary but not sufficient condition for diagnosing non-Gaussian entanglement.
Thus, based on three sets of standard form $V^n_1$, violations of criteria (\ref{eq4}) and (\ref{eq13}) imply \emph{fully inseparable} and \emph{genuine tripartite non-Gaussian entanglement}, respectively.
Besides, the elements in block $C_{k-lm}$ have state-independent properties \cite{triple.prl.2020}, which significantly simplify the complexity of experimental measurements compared to other cases where determining non-Gaussian entanglement requires two different measurement protocols, non-Gaussianity and inseparability \cite{jeong.np.8.564.2014,morin.np.8.570.2014,ra.np.non.2020}.

Those criteria based on three-mode correlations function composed of high-order creation or annihilation operators can also effectively determine the inseparability of TPS \cite{fei.pra.75.012311.2007}.
However, the eigenspectra of the observables in these criteria are discrete, so what they reveal is the high-order moments entanglement of Fock states.
The picture is especially evident when considering the two-mode squeezed vacuum, where the entanglement conditions $|\langle\hat{a}^n\hat{b}^n\rangle|>[\langle\hat{a}^{\dag n}\hat{a}^n\rangle\langle\hat{b}^{\dag n}\hat{b}^n\rangle]^{1/2}$ are always satisfied for any $n$ \cite{triple.prl.2020}.
This indicates that the high-order moments entanglement of Fock states revealed by these conditions is only 2nd-order moment entanglement --Gaussian entanglement-- from the perspective of continuous variables.
Therefore, the high-order moments in these criteria do not distinguish between Gaussian and non-Gaussian entanglement, which is clearly different from those in our criteria.

So far, there are other types of genuine tripartite entanglement criteria derived from the uncertainty principle \cite{shalm.np.9.1.2012, teh.pra.90.062337.2014,Shchukin.pra.90.012334.2014} and the Cauchy-Schwartz inequality \cite{agust.prl.125.020502}.
No assumptions were made there about the statistical properties of the states in deriving these criteria, so in principle they apply to any tripartite continuous-variable state, Gaussian or non-Gaussian.
In particular, these criteria are sufficient conditions for verifying entanglement rather than sufficient and necessary conditions.
When we claim that some kind of entanglement does not exist, we need to use necessary conditions of entanglement to support this conclusion, instead of sufficient conditions.
In other words, comparing two sufficient conditions for entanglements with two different states does not negate the existence of either Gaussian or non-Gaussian entanglements.
Therefore, what was proposed in \cite{agust.prl.125.020502} is a state-independent generalized entanglement condition, which cannot unambiguously reveal the novel notion of \emph{genuine tripartite non-Gaussian entanglement}.
In contrast, our criteria can capture tripartite non-Gaussian entanglement in moments of order $3n$.

To conclude this paper, we present the numerical verification of the proposed criteria.
\begin{figure}[htpb]
\centering
  \includegraphics[width=7cm]{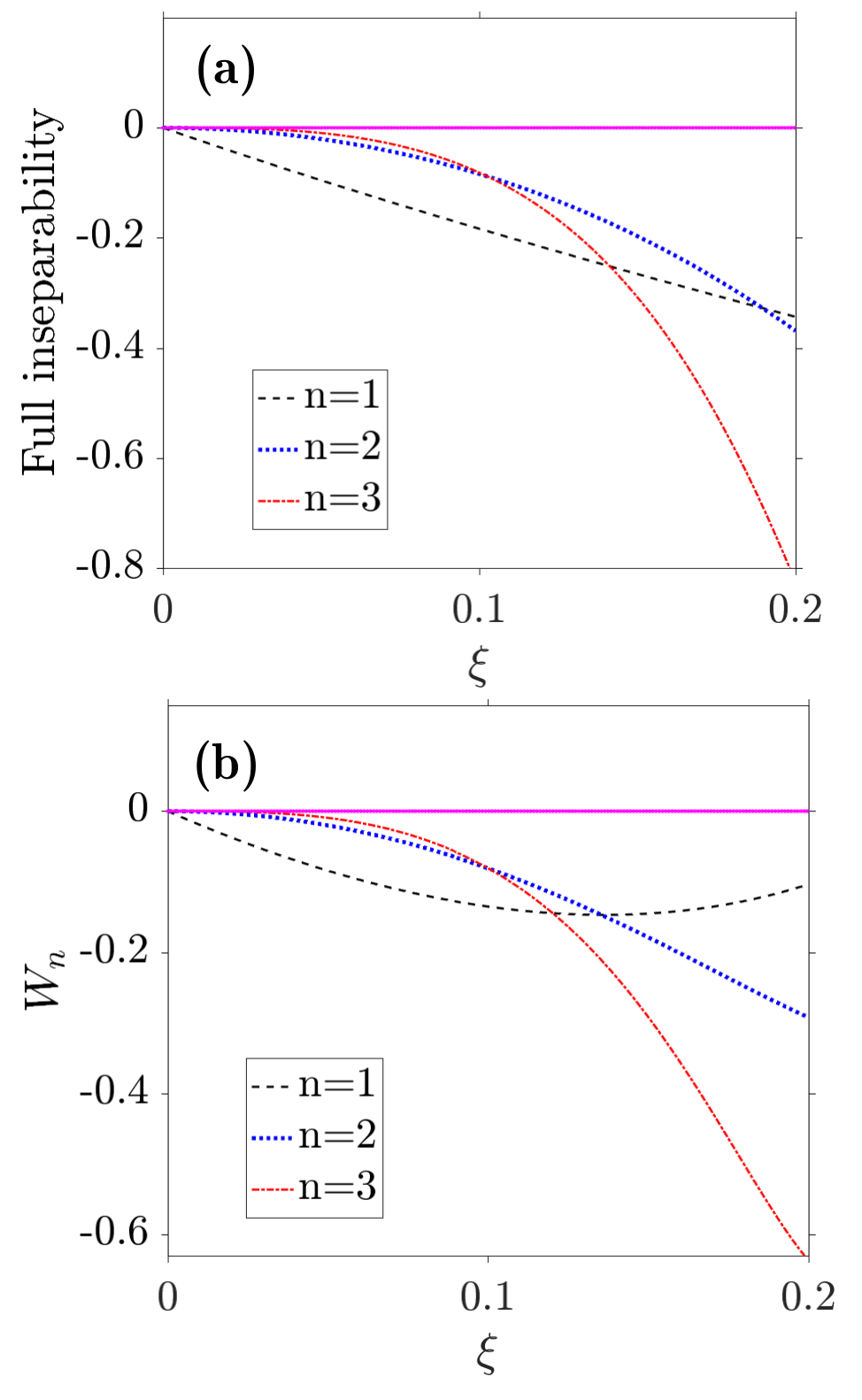}  
  \caption{Evolution of $F^n_1$ and $W_n$ as a function of the interaction strength $\xi$. $F^n_1<0$ and $W_n<0$ indicate full inseparability and genuine entanglement, respectively. $g_{1,n}=g_{2,n}=g_{3,n}=1$, $\xi=\kappa\alpha_p t$ and $\alpha_p=\sqrt{10}$. Note that $F^n_1=F^n_2=F^n_3$. }
  \label{fig1}
\end{figure}
Using the interaction Hamiltonian (\ref{eq1}), the master equations are solved numerically to deduce the final state of system at time $t$ considering that the initial state is vacuum for the triplets and a coherent mode $\alpha_p$ for the pump \cite{zhang.pra.013704.2021}.
Figure \ref{fig1}(a) shows the evolution of $F^n_1$ versus the interaction strength $\xi=\kappa\alpha_pt$, where $g_{1,n}=g_{2,n}=g_{3,n}=1$.
$F^n_1=F^n_2=F^n_3$ as expected from the symmetry of the TPS.
$F^{n=1,2,3}_1<0$ in the parameter region demonstrating full inseparability related to the 3rd-, 6th- and 9th-order covariance matrices of the TPS.
Figure \ref{fig1}(b) shows the evolution of $W_n$ versus the interaction strength.
$W_n < 0$ in the parameter region demonstrating genuine entanglement.
Starting from the vacuum, the genuine tripartite non-Gaussian entanglement is firstly loaded on the 3rd-order covariance matrices ($n=1$) and then gradually transitioned to the higher-order covariance matrices ($n=2,3$) with the increase of the interaction strength.
Therefore, TPS exhibit a hierarchical entanglement structure over a considerable range of parameters.

In summary, we proposed a hierarchy of sufficient and necessary conditions for the separability of higher-order moments for spontaneously-generated tripartite non-Gaussian states, which provides a systematic framework for characterizing non-Gaussian entanglement.
Besides, a hierarchy of genuine tripartite non-Gaussian entanglement criteria was proposed.
Compared with others, our proposal has the following advantages:
First, our criteria are more general since they can directly answer whether quantum states possess tripartite non-Gaussian entanglement, including full inseparability and genuine tripartite non-Gaussian entanglement.
Second, our strategy is platform-agnostic, as it works for any physical system as long as the information is encoded in continuous variables.
Third, our framework is naturally scalable and it constitutes a stepping stone to more sophisticated states and more than three parties.
Fourth, the physical quantities involved in our criteria are all experimentally accessible without quantum tomography.
Our proposal will have a significant impact on communities studying the theory of nonclassical correlations and exploring applications of non-Gaussian states in quantum information technology.
In particular, the hierarchy of entanglement structure of states may have special applications in quantum key distribution, quantum teleportation, or quantum computing, among others.
%

\section*{Acknowledgement}
This work was supported by the National Key Research and Development Program of China (2017YFA0303700), National Natural Science Foundation of China (61975159, 12204293) and by the Agence Nationale de la Recherche through Project TRIQUI (No. ANR 17-CE24-0041).


\begin{thebibliography}{51}%
\makeatletter
\providecommand \@ifxundefined [1]{%
 \@ifx{#1\undefined}
}%
\providecommand \@ifnum [1]{%
 \ifnum #1\expandafter \@firstoftwo
 \else \expandafter \@secondoftwo
 \fi
}%
\providecommand \@ifx [1]{%
 \ifx #1\expandafter \@firstoftwo
 \else \expandafter \@secondoftwo
 \fi
}%
\providecommand \natexlab [1]{#1}%
\providecommand \enquote  [1]{``#1''}%
\providecommand \bibnamefont  [1]{#1}%
\providecommand \bibfnamefont [1]{#1}%
\providecommand \citenamefont [1]{#1}%
\providecommand \href@noop [0]{\@secondoftwo}%
\providecommand \href [0]{\begingroup \@sanitize@url \@href}%
\providecommand \@href[1]{\@@startlink{#1}\@@href}%
\providecommand \@@href[1]{\endgroup#1\@@endlink}%
\providecommand \@sanitize@url [0]{\catcode `\\12\catcode `\$12\catcode
  `\&12\catcode `\#12\catcode `\^12\catcode `\_12\catcode `\%12\relax}%
\providecommand \@@startlink[1]{}%
\providecommand \@@endlink[0]{}%
\providecommand \url  [0]{\begingroup\@sanitize@url \@url }%
\providecommand \@url [1]{\endgroup\@href {#1}{\urlprefix }}%
\providecommand \urlprefix  [0]{URL }%
\providecommand \Eprint [0]{\href }%
\providecommand \doibase [0]{http://dx.doi.org/}%
\providecommand \selectlanguage [0]{\@gobble}%
\providecommand \bibinfo  [0]{\@secondoftwo}%
\providecommand \bibfield  [0]{\@secondoftwo}%
\providecommand \translation [1]{[#1]}%
\providecommand \BibitemOpen [0]{}%
\providecommand \bibitemStop [0]{}%
\providecommand \bibitemNoStop [0]{.\EOS\space}%
\providecommand \EOS [0]{\spacefactor3000\relax}%
\providecommand \BibitemShut  [1]{\csname bibitem#1\endcsname}%
\let\auto@bib@innerbib\@empty
\bibitem [{\citenamefont {Einstein}\ \emph {et~al.}(1935)\citenamefont
  {Einstein}, \citenamefont {Podolsky},\ and\ \citenamefont
  {Rosen}}]{einstein.pr.47.777.1935}%
  \BibitemOpen
  \bibfield  {author} {\bibinfo {author} {\bibfnamefont {A.}~\bibnamefont
  {Einstein}}, \bibinfo {author} {\bibfnamefont {B.}~\bibnamefont {Podolsky}},
  \ and\ \bibinfo {author} {\bibfnamefont {N.}~\bibnamefont {Rosen}},\ }\href
  {http://journals.aps.org/pr/abstract/10.1103/PhysRev.47.777} {\bibfield
  {journal} {\bibinfo  {journal} {Phys. Rev.}\ }\textbf {\bibinfo {volume}
  {47}},\ \bibinfo {pages} {777} (\bibinfo {year} {1935})}\BibitemShut
  {NoStop}%
\bibitem [{\citenamefont {Amico}\ \emph {et~al.}(2008)\citenamefont {Amico},
  \citenamefont {Fazio}, \citenamefont {Osterloh},\ and\ \citenamefont
  {Vedral}}]{Amico.rmp.80.517.2008}%
  \BibitemOpen
  \bibfield  {author} {\bibinfo {author} {\bibfnamefont {L.}~\bibnamefont
  {Amico}}, \bibinfo {author} {\bibfnamefont {R.}~\bibnamefont {Fazio}},
  \bibinfo {author} {\bibfnamefont {A.}~\bibnamefont {Osterloh}}, \ and\
  \bibinfo {author} {\bibfnamefont {V.}~\bibnamefont {Vedral}},\ }\href
  {\doibase 10.1103/RevModPhys.80.517} {\bibfield  {journal} {\bibinfo
  {journal} {Rev. Mod. Phys.}\ }\textbf {\bibinfo {volume} {80}},\ \bibinfo
  {pages} {517} (\bibinfo {year} {2008})}\BibitemShut {NoStop}%
\bibitem [{\citenamefont {Heidmann}\ \emph {et~al.}(1987)\citenamefont
  {Heidmann}, \citenamefont {Horowicz}, \citenamefont {Reynaud}, \citenamefont
  {Giacobino}, \citenamefont {Fabre},\ and\ \citenamefont
  {Camy}}]{Heidmann.prl.59.2555.1987}%
  \BibitemOpen
  \bibfield  {author} {\bibinfo {author} {\bibfnamefont {A.}~\bibnamefont
  {Heidmann}}, \bibinfo {author} {\bibfnamefont {R.~J.}\ \bibnamefont
  {Horowicz}}, \bibinfo {author} {\bibfnamefont {S.}~\bibnamefont {Reynaud}},
  \bibinfo {author} {\bibfnamefont {E.}~\bibnamefont {Giacobino}}, \bibinfo
  {author} {\bibfnamefont {C.}~\bibnamefont {Fabre}}, \ and\ \bibinfo {author}
  {\bibfnamefont {G.}~\bibnamefont {Camy}},\ }\href {\doibase
  10.1103/PhysRevLett.59.2555} {\bibfield  {journal} {\bibinfo  {journal}
  {Phys. Rev. Lett.}\ }\textbf {\bibinfo {volume} {59}},\ \bibinfo {pages}
  {2555} (\bibinfo {year} {1987})}\BibitemShut {NoStop}%
\bibitem [{\citenamefont {Zhang}\ \emph {et~al.}(2017)\citenamefont {Zhang},
  \citenamefont {Li}, \citenamefont {Zhang}, \citenamefont {Zhang},
  \citenamefont {Zhang},\ and\ \citenamefont
  {Xiao}}]{zhangda.pra.96.043847.2017}%
  \BibitemOpen
  \bibfield  {author} {\bibinfo {author} {\bibfnamefont {D.}~\bibnamefont
  {Zhang}}, \bibinfo {author} {\bibfnamefont {C.}~\bibnamefont {Li}}, \bibinfo
  {author} {\bibfnamefont {Z.}~\bibnamefont {Zhang}}, \bibinfo {author}
  {\bibfnamefont {Y.}~\bibnamefont {Zhang}}, \bibinfo {author} {\bibfnamefont
  {Y.}~\bibnamefont {Zhang}}, \ and\ \bibinfo {author} {\bibfnamefont
  {M.}~\bibnamefont {Xiao}},\ }\href {\doibase 10.1103/PhysRevA.96.043847}
  {\bibfield  {journal} {\bibinfo  {journal} {Phys. Rev. A}\ }\textbf {\bibinfo
  {volume} {96}},\ \bibinfo {pages} {043847} (\bibinfo {year}
  {2017})}\BibitemShut {NoStop}%
\bibitem [{\citenamefont {Yuen}(1976)}]{yuan.pra.13.2226.1976}%
  \BibitemOpen
  \bibfield  {author} {\bibinfo {author} {\bibfnamefont {H.~P.}\ \bibnamefont
  {Yuen}},\ }\href {\doibase 10.1103/PhysRevA.13.2226} {\bibfield  {journal}
  {\bibinfo  {journal} {Phys. Rev. A}\ }\textbf {\bibinfo {volume} {13}},\
  \bibinfo {pages} {2226} (\bibinfo {year} {1976})}\BibitemShut {NoStop}%
\bibitem [{\citenamefont {Schumaker}\ and\ \citenamefont
  {Caves}(1985)}]{caves.pra.31.3093.1985}%
  \BibitemOpen
  \bibfield  {author} {\bibinfo {author} {\bibfnamefont {B.~L.}\ \bibnamefont
  {Schumaker}}\ and\ \bibinfo {author} {\bibfnamefont {C.~M.}\ \bibnamefont
  {Caves}},\ }\href {\doibase 10.1103/PhysRevA.31.3093} {\bibfield  {journal}
  {\bibinfo  {journal} {Phys. Rev. A}\ }\textbf {\bibinfo {volume} {31}},\
  \bibinfo {pages} {3093} (\bibinfo {year} {1985})}\BibitemShut {NoStop}%
\bibitem [{\citenamefont {Scully}\ and\ \citenamefont
  {Zubairy}(1999)}]{scully.quantumoptics.1999}%
  \BibitemOpen
  \bibfield  {author} {\bibinfo {author} {\bibfnamefont {M.~O.}\ \bibnamefont
  {Scully}}\ and\ \bibinfo {author} {\bibfnamefont {M.~S.}\ \bibnamefont
  {Zubairy}},\ }\href@noop {} {\emph {\bibinfo {title} {Quantum optics}}}\
  (\bibinfo  {publisher} {American Association of Physics Teachers},\ \bibinfo
  {year} {1999})\BibitemShut {NoStop}%
\bibitem [{\citenamefont {Agarwal}(2012)}]{agarwal2012quantum}%
  \BibitemOpen
  \bibfield  {author} {\bibinfo {author} {\bibfnamefont {G.~S.}\ \bibnamefont
  {Agarwal}},\ }\href@noop {} {\emph {\bibinfo {title} {Quantum optics}}}\
  (\bibinfo  {publisher} {Cambridge University Press},\ \bibinfo {year}
  {2012})\BibitemShut {NoStop}%
\bibitem [{\citenamefont {Gerd}\ \emph {et~al.}(2007)\citenamefont {Gerd} \emph
  {et~al.}}]{gerd2007quantum}%
  \BibitemOpen
  \bibfield  {author} {\bibinfo {author} {\bibfnamefont {L.}~\bibnamefont
  {Gerd}} \emph {et~al.},\ }\href@noop {} {\emph {\bibinfo {title} {Quantum
  information with continuous variables of atoms and light}}}\ (\bibinfo
  {publisher} {World Scientific},\ \bibinfo {year} {2007})\BibitemShut
  {NoStop}%
\bibitem [{\citenamefont
  {Petz}(2007)}]{petz.quantuminformationprocessing.2007}%
  \BibitemOpen
  \bibfield  {author} {\bibinfo {author} {\bibfnamefont {D.}~\bibnamefont
  {Petz}},\ }\href@noop {} {\emph {\bibinfo {title} {Quantum information theory
  and quantum statistics}}}\ (\bibinfo  {publisher} {Springer Science \&
  Business Media},\ \bibinfo {year} {2007})\BibitemShut {NoStop}%
\bibitem [{\citenamefont {Braunstein}\ and\ \citenamefont
  {Pati}(2012)}]{braunstein2012quantum}%
  \BibitemOpen
  \bibfield  {author} {\bibinfo {author} {\bibfnamefont {S.~L.}\ \bibnamefont
  {Braunstein}}\ and\ \bibinfo {author} {\bibfnamefont {A.~K.}\ \bibnamefont
  {Pati}},\ }\href@noop {} {\emph {\bibinfo {title} {Quantum information with
  continuous variables}}}\ (\bibinfo  {publisher} {Springer Science \& Business
  Media},\ \bibinfo {year} {2012})\BibitemShut {NoStop}%
\bibitem [{\citenamefont {Braunstein}\ and\ \citenamefont {van
  Loock}(2005)}]{vanloock.rmp.77.513.2005}%
  \BibitemOpen
  \bibfield  {author} {\bibinfo {author} {\bibfnamefont {S.~L.}\ \bibnamefont
  {Braunstein}}\ and\ \bibinfo {author} {\bibfnamefont {P.}~\bibnamefont {van
  Loock}},\ }\href {\doibase 10.1103/RevModPhys.77.513} {\bibfield  {journal}
  {\bibinfo  {journal} {Rev. Mod. Phys.}\ }\textbf {\bibinfo {volume} {77}},\
  \bibinfo {pages} {513} (\bibinfo {year} {2005})}\BibitemShut {NoStop}%
\bibitem [{\citenamefont {Lloyd}\ and\ \citenamefont
  {Braunstein}(1999)}]{l1oyd.prl.82.1784.1999}%
  \BibitemOpen
  \bibfield  {author} {\bibinfo {author} {\bibfnamefont {S.}~\bibnamefont
  {Lloyd}}\ and\ \bibinfo {author} {\bibfnamefont {S.~L.}\ \bibnamefont
  {Braunstein}},\ }\href {\doibase 10.1103/PhysRevLett.82.1784} {\bibfield
  {journal} {\bibinfo  {journal} {Phys. Rev. Lett.}\ }\textbf {\bibinfo
  {volume} {82}},\ \bibinfo {pages} {1784} (\bibinfo {year}
  {1999})}\BibitemShut {NoStop}%
\bibitem [{\citenamefont {Bartlett}\ \emph {et~al.}(2002)\citenamefont
  {Bartlett}, \citenamefont {Sanders}, \citenamefont {Braunstein},\ and\
  \citenamefont {Nemoto}}]{bartlett.prl.88.097904.2002}%
  \BibitemOpen
  \bibfield  {author} {\bibinfo {author} {\bibfnamefont {S.~D.}\ \bibnamefont
  {Bartlett}}, \bibinfo {author} {\bibfnamefont {B.~C.}\ \bibnamefont
  {Sanders}}, \bibinfo {author} {\bibfnamefont {S.~L.}\ \bibnamefont
  {Braunstein}}, \ and\ \bibinfo {author} {\bibfnamefont {K.}~\bibnamefont
  {Nemoto}},\ }\href {\doibase 10.1103/PhysRevLett.88.097904} {\bibfield
  {journal} {\bibinfo  {journal} {Phys. Rev. Lett.}\ }\textbf {\bibinfo
  {volume} {88}},\ \bibinfo {pages} {097904} (\bibinfo {year}
  {2002})}\BibitemShut {NoStop}%
\bibitem [{\citenamefont {Menicucci}\ \emph {et~al.}(2006)\citenamefont
  {Menicucci}, \citenamefont {van Loock}, \citenamefont {Gu}, \citenamefont
  {Weedbrook}, \citenamefont {Ralph},\ and\ \citenamefont
  {Nielsen}}]{nielsen.prl.97.110501.2006}%
  \BibitemOpen
  \bibfield  {author} {\bibinfo {author} {\bibfnamefont {N.~C.}\ \bibnamefont
  {Menicucci}}, \bibinfo {author} {\bibfnamefont {P.}~\bibnamefont {van
  Loock}}, \bibinfo {author} {\bibfnamefont {M.}~\bibnamefont {Gu}}, \bibinfo
  {author} {\bibfnamefont {C.}~\bibnamefont {Weedbrook}}, \bibinfo {author}
  {\bibfnamefont {T.~C.}\ \bibnamefont {Ralph}}, \ and\ \bibinfo {author}
  {\bibfnamefont {M.~A.}\ \bibnamefont {Nielsen}},\ }\href {\doibase
  10.1103/PhysRevLett.97.110501} {\bibfield  {journal} {\bibinfo  {journal}
  {Phys. Rev. Lett.}\ }\textbf {\bibinfo {volume} {97}},\ \bibinfo {pages}
  {110501} (\bibinfo {year} {2006})}\BibitemShut {NoStop}%
\bibitem [{\citenamefont {Ohliger}\ \emph {et~al.}(2010)\citenamefont
  {Ohliger}, \citenamefont {Kieling},\ and\ \citenamefont
  {Eisert}}]{eisert.pra.82.042336.2010}%
  \BibitemOpen
  \bibfield  {author} {\bibinfo {author} {\bibfnamefont {M.}~\bibnamefont
  {Ohliger}}, \bibinfo {author} {\bibfnamefont {K.}~\bibnamefont {Kieling}}, \
  and\ \bibinfo {author} {\bibfnamefont {J.}~\bibnamefont {Eisert}},\ }\href
  {\doibase 10.1103/PhysRevA.82.042336} {\bibfield  {journal} {\bibinfo
  {journal} {Phys. Rev. A}\ }\textbf {\bibinfo {volume} {82}},\ \bibinfo
  {pages} {042336} (\bibinfo {year} {2010})}\BibitemShut {NoStop}%
\bibitem [{\citenamefont {Ohliger}\ and\ \citenamefont
  {Eisert}(2012)}]{eisert.pra.85.062318.2012}%
  \BibitemOpen
  \bibfield  {author} {\bibinfo {author} {\bibfnamefont {M.}~\bibnamefont
  {Ohliger}}\ and\ \bibinfo {author} {\bibfnamefont {J.}~\bibnamefont
  {Eisert}},\ }\href {\doibase 10.1103/PhysRevA.85.062318} {\bibfield
  {journal} {\bibinfo  {journal} {Phys. Rev. A}\ }\textbf {\bibinfo {volume}
  {85}},\ \bibinfo {pages} {062318} (\bibinfo {year} {2012})}\BibitemShut
  {NoStop}%
\bibitem [{\citenamefont {Hu}\ \emph {et~al.}(2020)\citenamefont {Hu},
  \citenamefont {Al-amri}, \citenamefont {Liao},\ and\ \citenamefont
  {Zubairy}}]{zubairy.pra.102.012608.2020}%
  \BibitemOpen
  \bibfield  {author} {\bibinfo {author} {\bibfnamefont {L.}~\bibnamefont
  {Hu}}, \bibinfo {author} {\bibfnamefont {M.}~\bibnamefont {Al-amri}},
  \bibinfo {author} {\bibfnamefont {Z.}~\bibnamefont {Liao}}, \ and\ \bibinfo
  {author} {\bibfnamefont {M.~S.}\ \bibnamefont {Zubairy}},\ }\href {\doibase
  10.1103/PhysRevA.102.012608} {\bibfield  {journal} {\bibinfo  {journal}
  {Phys. Rev. A}\ }\textbf {\bibinfo {volume} {102}},\ \bibinfo {pages}
  {012608} (\bibinfo {year} {2020})}\BibitemShut {NoStop}%
\bibitem [{\citenamefont {Opatrn\'y}\ \emph {et~al.}(2000)\citenamefont
  {Opatrn\'y}, \citenamefont {Kurizki},\ and\ \citenamefont
  {Welsch}}]{opatrn.pra.61.032302.2000}%
  \BibitemOpen
  \bibfield  {author} {\bibinfo {author} {\bibfnamefont {T.}~\bibnamefont
  {Opatrn\'y}}, \bibinfo {author} {\bibfnamefont {G.}~\bibnamefont {Kurizki}},
  \ and\ \bibinfo {author} {\bibfnamefont {D.-G.}\ \bibnamefont {Welsch}},\
  }\href {\doibase 10.1103/PhysRevA.61.032302} {\bibfield  {journal} {\bibinfo
  {journal} {Phys. Rev. A}\ }\textbf {\bibinfo {volume} {61}},\ \bibinfo
  {pages} {032302} (\bibinfo {year} {2000})}\BibitemShut {NoStop}%
\bibitem [{\citenamefont {Olivares}\ \emph {et~al.}(2003)\citenamefont
  {Olivares}, \citenamefont {Paris},\ and\ \citenamefont
  {Bonifacio}}]{olivares.pra.67.032314.2003}%
  \BibitemOpen
  \bibfield  {author} {\bibinfo {author} {\bibfnamefont {S.}~\bibnamefont
  {Olivares}}, \bibinfo {author} {\bibfnamefont {M.~G.~A.}\ \bibnamefont
  {Paris}}, \ and\ \bibinfo {author} {\bibfnamefont {R.}~\bibnamefont
  {Bonifacio}},\ }\href {\doibase 10.1103/PhysRevA.67.032314} {\bibfield
  {journal} {\bibinfo  {journal} {Phys. Rev. A}\ }\textbf {\bibinfo {volume}
  {67}},\ \bibinfo {pages} {032314} (\bibinfo {year} {2003})}\BibitemShut
  {NoStop}%
\bibitem [{\citenamefont {Dell'Anno}\ \emph {et~al.}(2007)\citenamefont
  {Dell'Anno}, \citenamefont {De~Siena}, \citenamefont {Albano},\ and\
  \citenamefont {Illuminati}}]{dell.pra.76.022301.2007}%
  \BibitemOpen
  \bibfield  {author} {\bibinfo {author} {\bibfnamefont {F.}~\bibnamefont
  {Dell'Anno}}, \bibinfo {author} {\bibfnamefont {S.}~\bibnamefont {De~Siena}},
  \bibinfo {author} {\bibfnamefont {L.}~\bibnamefont {Albano}}, \ and\ \bibinfo
  {author} {\bibfnamefont {F.}~\bibnamefont {Illuminati}},\ }\href {\doibase
  10.1103/PhysRevA.76.022301} {\bibfield  {journal} {\bibinfo  {journal} {Phys.
  Rev. A}\ }\textbf {\bibinfo {volume} {76}},\ \bibinfo {pages} {022301}
  (\bibinfo {year} {2007})}\BibitemShut {NoStop}%
\bibitem [{\citenamefont {Strobel}\ \emph {et~al.}(2014)\citenamefont
  {Strobel}, \citenamefont {Muessel}, \citenamefont {Linnemann}, \citenamefont
  {Zibold}, \citenamefont {Hume}, \citenamefont {Pezz{\`e}}, \citenamefont
  {Smerzi},\ and\ \citenamefont {Oberthaler}}]{augusto.science.345.2014}%
  \BibitemOpen
  \bibfield  {author} {\bibinfo {author} {\bibfnamefont {H.}~\bibnamefont
  {Strobel}}, \bibinfo {author} {\bibfnamefont {W.}~\bibnamefont {Muessel}},
  \bibinfo {author} {\bibfnamefont {D.}~\bibnamefont {Linnemann}}, \bibinfo
  {author} {\bibfnamefont {T.}~\bibnamefont {Zibold}}, \bibinfo {author}
  {\bibfnamefont {D.~B.}\ \bibnamefont {Hume}}, \bibinfo {author}
  {\bibfnamefont {L.}~\bibnamefont {Pezz{\`e}}}, \bibinfo {author}
  {\bibfnamefont {A.}~\bibnamefont {Smerzi}}, \ and\ \bibinfo {author}
  {\bibfnamefont {M.~K.}\ \bibnamefont {Oberthaler}},\ }\href
  {https://science.sciencemag.org/content/345/6195/424} {\bibfield  {journal}
  {\bibinfo  {journal} {Science}\ }\textbf {\bibinfo {volume} {345}},\ \bibinfo
  {pages} {424} (\bibinfo {year} {2014})}\BibitemShut {NoStop}%
\bibitem [{\citenamefont {Chabaud}\ and\ \citenamefont
  {Walschaers}(2022)}]{Mattia2022resources}%
  \BibitemOpen
  \bibfield  {author} {\bibinfo {author} {\bibfnamefont {U.}~\bibnamefont
  {Chabaud}}\ and\ \bibinfo {author} {\bibfnamefont {M.}~\bibnamefont
  {Walschaers}},\ }\href {https://arxiv.org/abs/2207.11781} {\bibfield
  {journal} {\bibinfo  {journal} {arXiv preprint arXiv:2207.11781}\ } (\bibinfo
  {year} {2022})}\BibitemShut {NoStop}%
\bibitem [{\citenamefont {Jeong}\ \emph {et~al.}(2014)\citenamefont {Jeong},
  \citenamefont {Zavatta}, \citenamefont {Kang}, \citenamefont {Lee},
  \citenamefont {Costanzo}, \citenamefont {Grandi}, \citenamefont {Ralph},\
  and\ \citenamefont {Bellini}}]{jeong.np.8.564.2014}%
  \BibitemOpen
  \bibfield  {author} {\bibinfo {author} {\bibfnamefont {H.}~\bibnamefont
  {Jeong}}, \bibinfo {author} {\bibfnamefont {A.}~\bibnamefont {Zavatta}},
  \bibinfo {author} {\bibfnamefont {M.}~\bibnamefont {Kang}}, \bibinfo {author}
  {\bibfnamefont {S.-W.}\ \bibnamefont {Lee}}, \bibinfo {author} {\bibfnamefont
  {L.~S.}\ \bibnamefont {Costanzo}}, \bibinfo {author} {\bibfnamefont
  {S.}~\bibnamefont {Grandi}}, \bibinfo {author} {\bibfnamefont {T.~C.}\
  \bibnamefont {Ralph}}, \ and\ \bibinfo {author} {\bibfnamefont
  {M.}~\bibnamefont {Bellini}},\ }\href
  {https://www.nature.com/articles/nphoton.2014.136} {\bibfield  {journal}
  {\bibinfo  {journal} {Nat. Photonics}\ }\textbf {\bibinfo {volume} {8}},\
  \bibinfo {pages} {564} (\bibinfo {year} {2014})}\BibitemShut {NoStop}%
\bibitem [{\citenamefont {Morin}\ \emph {et~al.}(2014)\citenamefont {Morin},
  \citenamefont {Huang}, \citenamefont {Liu}, \citenamefont {Le~Jeannic},
  \citenamefont {Fabre},\ and\ \citenamefont {Laurat}}]{morin.np.8.570.2014}%
  \BibitemOpen
  \bibfield  {author} {\bibinfo {author} {\bibfnamefont {O.}~\bibnamefont
  {Morin}}, \bibinfo {author} {\bibfnamefont {K.}~\bibnamefont {Huang}},
  \bibinfo {author} {\bibfnamefont {J.}~\bibnamefont {Liu}}, \bibinfo {author}
  {\bibfnamefont {H.}~\bibnamefont {Le~Jeannic}}, \bibinfo {author}
  {\bibfnamefont {C.}~\bibnamefont {Fabre}}, \ and\ \bibinfo {author}
  {\bibfnamefont {J.}~\bibnamefont {Laurat}},\ }\href
  {https://www.nature.com/articles/nphoton.2014.137} {\bibfield  {journal}
  {\bibinfo  {journal} {Nat. Photonics}\ }\textbf {\bibinfo {volume} {8}},\
  \bibinfo {pages} {570} (\bibinfo {year} {2014})}\BibitemShut {NoStop}%
\bibitem [{\citenamefont {Ra}\ \emph {et~al.}(2020)\citenamefont {Ra},
  \citenamefont {Dufour}, \citenamefont {Walschaers}, \citenamefont {Jacquard},
  \citenamefont {Michel}, \citenamefont {Fabre},\ and\ \citenamefont
  {Treps}}]{ra.np.non.2020}%
  \BibitemOpen
  \bibfield  {author} {\bibinfo {author} {\bibfnamefont {Y.-S.}\ \bibnamefont
  {Ra}}, \bibinfo {author} {\bibfnamefont {A.}~\bibnamefont {Dufour}}, \bibinfo
  {author} {\bibfnamefont {M.}~\bibnamefont {Walschaers}}, \bibinfo {author}
  {\bibfnamefont {C.}~\bibnamefont {Jacquard}}, \bibinfo {author}
  {\bibfnamefont {T.}~\bibnamefont {Michel}}, \bibinfo {author} {\bibfnamefont
  {C.}~\bibnamefont {Fabre}}, \ and\ \bibinfo {author} {\bibfnamefont
  {N.}~\bibnamefont {Treps}},\ }\href
  {https://www.nature.com/articles/s41567-019-0726-y} {\bibfield  {journal}
  {\bibinfo  {journal} {Nat. Phys.}\ }\textbf {\bibinfo {volume} {16}},\
  \bibinfo {pages} {144} (\bibinfo {year} {2020})}\BibitemShut {NoStop}%
\bibitem [{\citenamefont {Chang}\ \emph {et~al.}(2020)\citenamefont {Chang},
  \citenamefont {Sab\'{\i}n}, \citenamefont {Forn-D\'{\i}az}, \citenamefont
  {Quijandr\'{\i}a}, \citenamefont {Vadiraj}, \citenamefont {Nsanzineza},
  \citenamefont {Johansson},\ and\ \citenamefont
  {Wilson}}]{chang.prx.10.011011.2020}%
  \BibitemOpen
  \bibfield  {author} {\bibinfo {author} {\bibfnamefont {C.~W.~S.}\
  \bibnamefont {Chang}}, \bibinfo {author} {\bibfnamefont {C.}~\bibnamefont
  {Sab\'{\i}n}}, \bibinfo {author} {\bibfnamefont {P.}~\bibnamefont
  {Forn-D\'{\i}az}}, \bibinfo {author} {\bibfnamefont {F.}~\bibnamefont
  {Quijandr\'{\i}a}}, \bibinfo {author} {\bibfnamefont {A.~M.}\ \bibnamefont
  {Vadiraj}}, \bibinfo {author} {\bibfnamefont {I.}~\bibnamefont {Nsanzineza}},
  \bibinfo {author} {\bibfnamefont {G.}~\bibnamefont {Johansson}}, \ and\
  \bibinfo {author} {\bibfnamefont {C.~M.}\ \bibnamefont {Wilson}},\ }\href
  {\doibase 10.1103/PhysRevX.10.011011} {\bibfield  {journal} {\bibinfo
  {journal} {Phys. Rev. X}\ }\textbf {\bibinfo {volume} {10}},\ \bibinfo
  {pages} {011011} (\bibinfo {year} {2020})}\BibitemShut {NoStop}%
\bibitem [{\citenamefont {Moebius}\ \emph {et~al.}(2016)\citenamefont
  {Moebius}, \citenamefont {Herrera}, \citenamefont {Griesse-Nascimento},
  \citenamefont {Reshef}, \citenamefont {Evans}, \citenamefont {Guerreschi},
  \citenamefont {Aspuru-Guzik},\ and\ \citenamefont
  {Mazur}}]{Moebius.oe.9.9932.2016}%
  \BibitemOpen
  \bibfield  {author} {\bibinfo {author} {\bibfnamefont {M.~G.}\ \bibnamefont
  {Moebius}}, \bibinfo {author} {\bibfnamefont {F.}~\bibnamefont {Herrera}},
  \bibinfo {author} {\bibfnamefont {S.}~\bibnamefont {Griesse-Nascimento}},
  \bibinfo {author} {\bibfnamefont {O.}~\bibnamefont {Reshef}}, \bibinfo
  {author} {\bibfnamefont {C.~C.}\ \bibnamefont {Evans}}, \bibinfo {author}
  {\bibfnamefont {G.~G.}\ \bibnamefont {Guerreschi}}, \bibinfo {author}
  {\bibfnamefont {A.}~\bibnamefont {Aspuru-Guzik}}, \ and\ \bibinfo {author}
  {\bibfnamefont {E.}~\bibnamefont {Mazur}},\ }\href
  {http://www.opticsexpress.org/abstract.cfm?URI=oe-24-9-9932} {\bibfield
  {journal} {\bibinfo  {journal} {Opt. Express}\ }\textbf {\bibinfo {volume}
  {24}},\ \bibinfo {pages} {9932} (\bibinfo {year} {2016})}\BibitemShut
  {NoStop}%
\bibitem [{\citenamefont {Douady}\ and\ \citenamefont
  {Boulanger}(2004)}]{douady.23.2794.ol.2004}%
  \BibitemOpen
  \bibfield  {author} {\bibinfo {author} {\bibfnamefont {J.}~\bibnamefont
  {Douady}}\ and\ \bibinfo {author} {\bibfnamefont {B.}~\bibnamefont
  {Boulanger}},\ }\href {\doibase 10.1364/OL.29.002794} {\bibfield  {journal}
  {\bibinfo  {journal} {Opt. Lett.}\ }\textbf {\bibinfo {volume} {29}},\
  \bibinfo {pages} {2794} (\bibinfo {year} {2004})}\BibitemShut {NoStop}%
\bibitem [{\citenamefont {Cavanna}\ \emph {et~al.}(2020)\citenamefont
  {Cavanna}, \citenamefont {Hammer}, \citenamefont {Okoth}, \citenamefont
  {Ortiz-Ricardo}, \citenamefont {Cruz-Ramirez}, \citenamefont {Garay-Palmett},
  \citenamefont {U'Ren}, \citenamefont {Frosz}, \citenamefont {Jiang},
  \citenamefont {Joly},\ and\ \citenamefont
  {Chekhova}}]{cavanna.pra.101.033840.2020}%
  \BibitemOpen
  \bibfield  {author} {\bibinfo {author} {\bibfnamefont {A.}~\bibnamefont
  {Cavanna}}, \bibinfo {author} {\bibfnamefont {J.}~\bibnamefont {Hammer}},
  \bibinfo {author} {\bibfnamefont {C.}~\bibnamefont {Okoth}}, \bibinfo
  {author} {\bibfnamefont {E.}~\bibnamefont {Ortiz-Ricardo}}, \bibinfo {author}
  {\bibfnamefont {H.}~\bibnamefont {Cruz-Ramirez}}, \bibinfo {author}
  {\bibfnamefont {K.}~\bibnamefont {Garay-Palmett}}, \bibinfo {author}
  {\bibfnamefont {A.~B.}\ \bibnamefont {U'Ren}}, \bibinfo {author}
  {\bibfnamefont {M.~H.}\ \bibnamefont {Frosz}}, \bibinfo {author}
  {\bibfnamefont {X.}~\bibnamefont {Jiang}}, \bibinfo {author} {\bibfnamefont
  {N.~Y.}\ \bibnamefont {Joly}}, \ and\ \bibinfo {author} {\bibfnamefont
  {M.~V.}\ \bibnamefont {Chekhova}},\ }\href {\doibase
  10.1103/PhysRevA.101.033840} {\bibfield  {journal} {\bibinfo  {journal}
  {Phys. Rev. A}\ }\textbf {\bibinfo {volume} {101}},\ \bibinfo {pages}
  {033840} (\bibinfo {year} {2020})}\BibitemShut {NoStop}%
\bibitem [{\citenamefont {Li}\ \emph {et~al.}(2020)\citenamefont {Li},
  \citenamefont {Cai}, \citenamefont {Wu}, \citenamefont {Liu}, \citenamefont
  {Xiong}, \citenamefont {Li},\ and\ \citenamefont
  {Zhang}}]{kangkang.aqt.35.2020}%
  \BibitemOpen
  \bibfield  {author} {\bibinfo {author} {\bibfnamefont {K.}~\bibnamefont
  {Li}}, \bibinfo {author} {\bibfnamefont {Y.}~\bibnamefont {Cai}}, \bibinfo
  {author} {\bibfnamefont {J.}~\bibnamefont {Wu}}, \bibinfo {author}
  {\bibfnamefont {Y.}~\bibnamefont {Liu}}, \bibinfo {author} {\bibfnamefont
  {S.}~\bibnamefont {Xiong}}, \bibinfo {author} {\bibfnamefont
  {Y.}~\bibnamefont {Li}}, \ and\ \bibinfo {author} {\bibfnamefont
  {Y.}~\bibnamefont {Zhang}},\ }\href {\doibase
  https://doi.org/10.1002/qute.201900119} {\bibfield  {journal} {\bibinfo
  {journal} {Adv. Quantum Technol.}\ }\textbf {\bibinfo {volume} {3}},\
  \bibinfo {pages} {1900119} (\bibinfo {year} {2020})}\BibitemShut {NoStop}%
\bibitem [{\citenamefont {Werner}\ and\ \citenamefont
  {Wolf}(2001)}]{werner.prl.86.3658.2001}%
  \BibitemOpen
  \bibfield  {author} {\bibinfo {author} {\bibfnamefont {R.~F.}\ \bibnamefont
  {Werner}}\ and\ \bibinfo {author} {\bibfnamefont {M.~M.}\ \bibnamefont
  {Wolf}},\ }\href {\doibase 10.1103/PhysRevLett.86.3658} {\bibfield  {journal}
  {\bibinfo  {journal} {Phys. Rev. Lett.}\ }\textbf {\bibinfo {volume} {86}},\
  \bibinfo {pages} {3658} (\bibinfo {year} {2001})}\BibitemShut {NoStop}%
\bibitem [{\citenamefont {van Loock}\ and\ \citenamefont
  {Furusawa}(2003)}]{vanlook.pra.67.052315.2003}%
  \BibitemOpen
  \bibfield  {author} {\bibinfo {author} {\bibfnamefont {P.}~\bibnamefont {van
  Loock}}\ and\ \bibinfo {author} {\bibfnamefont {A.}~\bibnamefont
  {Furusawa}},\ }\href {\doibase 10.1103/PhysRevA.67.052315} {\bibfield
  {journal} {\bibinfo  {journal} {Phys. Rev. A}\ }\textbf {\bibinfo {volume}
  {67}},\ \bibinfo {pages} {052315} (\bibinfo {year} {2003})}\BibitemShut
  {NoStop}%
\bibitem [{\citenamefont {Gonz\'alez}\ \emph {et~al.}(2018)\citenamefont
  {Gonz\'alez}, \citenamefont {Borne}, \citenamefont {Boulanger}, \citenamefont
  {Levenson},\ and\ \citenamefont {Bencheikh}}]{kamel.prl.120.043601.2018}%
  \BibitemOpen
  \bibfield  {author} {\bibinfo {author} {\bibfnamefont {E.~A.~R.}\
  \bibnamefont {Gonz\'alez}}, \bibinfo {author} {\bibfnamefont
  {A.}~\bibnamefont {Borne}}, \bibinfo {author} {\bibfnamefont
  {B.}~\bibnamefont {Boulanger}}, \bibinfo {author} {\bibfnamefont {J.~A.}\
  \bibnamefont {Levenson}}, \ and\ \bibinfo {author} {\bibfnamefont
  {K.}~\bibnamefont {Bencheikh}},\ }\href {\doibase
  10.1103/PhysRevLett.120.043601} {\bibfield  {journal} {\bibinfo  {journal}
  {Phys. Rev. Lett.}\ }\textbf {\bibinfo {volume} {120}},\ \bibinfo {pages}
  {043601} (\bibinfo {year} {2018})}\BibitemShut {NoStop}%
\bibitem [{\citenamefont {Agust\'{\i}}\ \emph {et~al.}(2020)\citenamefont
  {Agust\'{\i}}, \citenamefont {Chang}, \citenamefont {Quijandr\'{\i}a},
  \citenamefont {Johansson}, \citenamefont {Wilson},\ and\ \citenamefont
  {Sab\'{\i}n}}]{agust.prl.125.020502}%
  \BibitemOpen
  \bibfield  {author} {\bibinfo {author} {\bibfnamefont {A.}~\bibnamefont
  {Agust\'{\i}}}, \bibinfo {author} {\bibfnamefont {C.~W.~S.}\ \bibnamefont
  {Chang}}, \bibinfo {author} {\bibfnamefont {F.}~\bibnamefont
  {Quijandr\'{\i}a}}, \bibinfo {author} {\bibfnamefont {G.}~\bibnamefont
  {Johansson}}, \bibinfo {author} {\bibfnamefont {C.~M.}\ \bibnamefont
  {Wilson}}, \ and\ \bibinfo {author} {\bibfnamefont {C.}~\bibnamefont
  {Sab\'{\i}n}},\ }\href {\doibase 10.1103/PhysRevLett.125.020502} {\bibfield
  {journal} {\bibinfo  {journal} {Phys. Rev. Lett.}\ }\textbf {\bibinfo
  {volume} {125}},\ \bibinfo {pages} {020502} (\bibinfo {year}
  {2020})}\BibitemShut {NoStop}%
\bibitem [{\citenamefont {Zhang}\ \emph {et~al.}(2021)\citenamefont {Zhang},
  \citenamefont {Cai}, \citenamefont {Zheng}, \citenamefont {Barral},
  \citenamefont {Zhang}, \citenamefont {Xiao},\ and\ \citenamefont
  {Bencheikh}}]{zhang.pra.013704.2021}%
  \BibitemOpen
  \bibfield  {author} {\bibinfo {author} {\bibfnamefont {D.}~\bibnamefont
  {Zhang}}, \bibinfo {author} {\bibfnamefont {Y.}~\bibnamefont {Cai}}, \bibinfo
  {author} {\bibfnamefont {Z.}~\bibnamefont {Zheng}}, \bibinfo {author}
  {\bibfnamefont {D.}~\bibnamefont {Barral}}, \bibinfo {author} {\bibfnamefont
  {Y.}~\bibnamefont {Zhang}}, \bibinfo {author} {\bibfnamefont
  {M.}~\bibnamefont {Xiao}}, \ and\ \bibinfo {author} {\bibfnamefont
  {K.}~\bibnamefont {Bencheikh}},\ }\href {\doibase
  10.1103/PhysRevA.103.013704} {\bibfield  {journal} {\bibinfo  {journal}
  {Phys. Rev. A}\ }\textbf {\bibinfo {volume} {103}},\ \bibinfo {pages}
  {013704} (\bibinfo {year} {2021})}\BibitemShut {NoStop}%
\bibitem [{\citenamefont {Hillery}\ and\ \citenamefont
  {Zubairy}(2006)}]{mm.prl.96.050503.2006}%
  \BibitemOpen
  \bibfield  {author} {\bibinfo {author} {\bibfnamefont {M.}~\bibnamefont
  {Hillery}}\ and\ \bibinfo {author} {\bibfnamefont {M.~S.}\ \bibnamefont
  {Zubairy}},\ }\href {\doibase 10.1103/PhysRevLett.96.050503} {\bibfield
  {journal} {\bibinfo  {journal} {Phys. Rev. Lett.}\ }\textbf {\bibinfo
  {volume} {96}},\ \bibinfo {pages} {050503} (\bibinfo {year}
  {2006})}\BibitemShut {NoStop}%
\bibitem [{\citenamefont {Miranowicz}\ \emph {et~al.}(2009)\citenamefont
  {Miranowicz}, \citenamefont {Piani}, \citenamefont {Horodecki},\ and\
  \citenamefont {Horodecki}}]{adam.pra.80.0523032009}%
  \BibitemOpen
  \bibfield  {author} {\bibinfo {author} {\bibfnamefont {A.}~\bibnamefont
  {Miranowicz}}, \bibinfo {author} {\bibfnamefont {M.}~\bibnamefont {Piani}},
  \bibinfo {author} {\bibfnamefont {P.}~\bibnamefont {Horodecki}}, \ and\
  \bibinfo {author} {\bibfnamefont {R.}~\bibnamefont {Horodecki}},\ }\href
  {\doibase 10.1103/PhysRevA.80.052303} {\bibfield  {journal} {\bibinfo
  {journal} {Phys. Rev. A}\ }\textbf {\bibinfo {volume} {80}},\ \bibinfo
  {pages} {052303} (\bibinfo {year} {2009})}\BibitemShut {NoStop}%
\bibitem [{\citenamefont {Hillery}\ \emph {et~al.}(2010)\citenamefont
  {Hillery}, \citenamefont {Dung},\ and\ \citenamefont
  {Zheng}}]{hillery.pra.81.062322.2010}%
  \BibitemOpen
  \bibfield  {author} {\bibinfo {author} {\bibfnamefont {M.}~\bibnamefont
  {Hillery}}, \bibinfo {author} {\bibfnamefont {H.~T.}\ \bibnamefont {Dung}}, \
  and\ \bibinfo {author} {\bibfnamefont {H.}~\bibnamefont {Zheng}},\ }\href
  {\doibase 10.1103/PhysRevA.81.062322} {\bibfield  {journal} {\bibinfo
  {journal} {Phys. Rev. A}\ }\textbf {\bibinfo {volume} {81}},\ \bibinfo
  {pages} {062322} (\bibinfo {year} {2010})}\BibitemShut {NoStop}%
\bibitem [{\citenamefont {Walschaers}(2021)}]{Mattia.PRXQuantum.2.030204.2021}%
  \BibitemOpen
  \bibfield  {author} {\bibinfo {author} {\bibfnamefont {M.}~\bibnamefont
  {Walschaers}},\ }\href {\doibase 10.1103/PRXQuantum.2.030204} {\bibfield
  {journal} {\bibinfo  {journal} {PRX Quantum}\ }\textbf {\bibinfo {volume}
  {2}},\ \bibinfo {pages} {030204} (\bibinfo {year} {2021})}\BibitemShut
  {NoStop}%
\bibitem [{\citenamefont {Tian}\ \emph {et~al.}(2022)\citenamefont {Tian},
  \citenamefont {Xiang}, \citenamefont {Sun}, \citenamefont {Fadel},\ and\
  \citenamefont {He}}]{tian.prapplied.18.024065.2022}%
  \BibitemOpen
  \bibfield  {author} {\bibinfo {author} {\bibfnamefont {M.}~\bibnamefont
  {Tian}}, \bibinfo {author} {\bibfnamefont {Y.}~\bibnamefont {Xiang}},
  \bibinfo {author} {\bibfnamefont {F.-X.}\ \bibnamefont {Sun}}, \bibinfo
  {author} {\bibfnamefont {M.}~\bibnamefont {Fadel}}, \ and\ \bibinfo {author}
  {\bibfnamefont {Q.}~\bibnamefont {He}},\ }\href {\doibase
  10.1103/PhysRevApplied.18.024065} {\bibfield  {journal} {\bibinfo  {journal}
  {Phys. Rev. Applied}\ }\textbf {\bibinfo {volume} {18}},\ \bibinfo {pages}
  {024065} (\bibinfo {year} {2022})}\BibitemShut {NoStop}%
\bibitem [{\citenamefont {at~http://link.aps.org/supplemental/XXXXXX for
  details on the commutation relations of operators}\ and\
  \citenamefont {the proof of~two lemmas}()}]{triple.prl.2020}%
  \BibitemOpen
  \bibfield  {author} {\bibinfo {author} {\bibfnamefont {See Supplemental Material at}\
  \bibnamefont {http://link.aps.org/ supplemental/XXXXXX for details on the
  commutation relations of operators,}}\ \ \bibinfo
  {author} {\bibnamefont {the basic properties of TPS, the derivation processes of standard form $V^n_1$ and genuine tripartite entanglement criterion}}}\href
  {http://link.aps.org/supplemental/XXXXXXXXXXXX} {\ }\BibitemShut {NoStop}
\bibitem [{\citenamefont {Shchukin}\ and\ \citenamefont {van
  Loock}(2016)}]{shchukin.pra.93.032114.2016}%
  \BibitemOpen
  \bibfield  {author} {\bibinfo {author} {\bibfnamefont {E.}~\bibnamefont
  {Shchukin}}\ and\ \bibinfo {author} {\bibfnamefont {P.}~\bibnamefont {van
  Loock}},\ }\href {\doibase 10.1103/PhysRevA.93.032114} {\bibfield  {journal}
  {\bibinfo  {journal} {Phys. Rev. A}\ }\textbf {\bibinfo {volume} {93}},\
  \bibinfo {pages} {032114} (\bibinfo {year} {2016})}\BibitemShut {NoStop}%
\bibitem [{\citenamefont {Shchukin}\ and\ \citenamefont {van
  Loock}(2014)}]{Shchukin.pra.90.012334.2014}%
  \BibitemOpen
  \bibfield  {author} {\bibinfo {author} {\bibfnamefont {E.}~\bibnamefont
  {Shchukin}}\ and\ \bibinfo {author} {\bibfnamefont {P.}~\bibnamefont {van
  Loock}},\ }\href {\doibase 10.1103/PhysRevA.90.012334} {\bibfield  {journal}
  {\bibinfo  {journal} {Phys. Rev. A}\ }\textbf {\bibinfo {volume} {90}},\
  \bibinfo {pages} {012334} (\bibinfo {year} {2014})}\BibitemShut {NoStop}%
\bibitem [{\citenamefont {Horn}\ and\ \citenamefont
  {Johnson}(2012)}]{horn2012matrix}%
  \BibitemOpen
  \bibfield  {author} {\bibinfo {author} {\bibfnamefont {R.~A.}\ \bibnamefont
  {Horn}}\ and\ \bibinfo {author} {\bibfnamefont {C.~R.}\ \bibnamefont
  {Johnson}},\ }\href@noop {} {\emph {\bibinfo {title} {Matrix analysis}}}\
  (\bibinfo  {publisher} {Cambridge university press},\ \bibinfo {year}
  {2012})\BibitemShut {NoStop}%
\bibitem [{\citenamefont {Simon}(2000)}]{simon.prl.84.2726.2000}%
  \BibitemOpen
  \bibfield  {author} {\bibinfo {author} {\bibfnamefont {R.}~\bibnamefont
  {Simon}},\ }\href {\doibase 10.1103/PhysRevLett.84.2726} {\bibfield
  {journal} {\bibinfo  {journal} {Phys. Rev. Lett.}\ }\textbf {\bibinfo
  {volume} {84}},\ \bibinfo {pages} {2726} (\bibinfo {year}
  {2000})}\BibitemShut {NoStop}%
\bibitem [{\citenamefont {Duan}\ \emph {et~al.}(2000)\citenamefont {Duan},
  \citenamefont {Giedke}, \citenamefont {Cirac},\ and\ \citenamefont
  {Zoller}}]{duan.prl.84.2722.2000}%
  \BibitemOpen
  \bibfield  {author} {\bibinfo {author} {\bibfnamefont {L.-M.}\ \bibnamefont
  {Duan}}, \bibinfo {author} {\bibfnamefont {G.}~\bibnamefont {Giedke}},
  \bibinfo {author} {\bibfnamefont {J.~I.}\ \bibnamefont {Cirac}}, \ and\
  \bibinfo {author} {\bibfnamefont {P.}~\bibnamefont {Zoller}},\ }\href
  {\doibase 10.1103/PhysRevLett.84.2722} {\bibfield  {journal} {\bibinfo
  {journal} {Phys. Rev. Lett.}\ }\textbf {\bibinfo {volume} {84}},\ \bibinfo
  {pages} {2722} (\bibinfo {year} {2000})}\BibitemShut {NoStop}%
\bibitem [{\citenamefont {Hyllus}\ and\ \citenamefont
  {Eisert}(2006)}]{eisert.njp.8.51.2006}%
  \BibitemOpen
  \bibfield  {author} {\bibinfo {author} {\bibfnamefont {P.}~\bibnamefont
  {Hyllus}}\ and\ \bibinfo {author} {\bibfnamefont {J.}~\bibnamefont
  {Eisert}},\ }\href {\doibase 10.1088/1367-2630/8/4/051} {\bibfield  {journal}
  {\bibinfo  {journal} {New J. Phys.}\ }\textbf {\bibinfo {volume} {8}},\
  \bibinfo {pages} {51} (\bibinfo {year} {2006})}\BibitemShut {NoStop}%
\bibitem [{\citenamefont {Shalm}\ \emph {et~al.}(2012)\citenamefont {Shalm},
  \citenamefont {Hamel}, \citenamefont {Yan}, \citenamefont {Simon},\ and\
  \citenamefont {Jennewein}}]{shalm.np.9.1.2012}%
  \BibitemOpen
  \bibfield  {author} {\bibinfo {author} {\bibfnamefont {L.~K.}\ \bibnamefont
  {Shalm}}, \bibinfo {author} {\bibfnamefont {D.~R.}\ \bibnamefont {Hamel}},
  \bibinfo {author} {\bibfnamefont {Z.}~\bibnamefont {Yan}}, \bibinfo {author}
  {\bibfnamefont {C.}~\bibnamefont {Simon}}, \ and\ \bibinfo {author}
  {\bibfnamefont {T.}~\bibnamefont {Jennewein}},\ }\href
  {https://www.nature.com/articles/nphys2492} {\bibfield  {journal} {\bibinfo
  {journal} {Nat. Phys.}\ }\textbf {\bibinfo {volume} {9}},\ \bibinfo {pages}
  {19} (\bibinfo {year} {2012})}\BibitemShut {NoStop}%
\bibitem [{\citenamefont {Teh}\ and\ \citenamefont
  {Reid}(2014)}]{teh.pra.90.062337.2014}%
  \BibitemOpen
  \bibfield  {author} {\bibinfo {author} {\bibfnamefont {R.~Y.}\ \bibnamefont
  {Teh}}\ and\ \bibinfo {author} {\bibfnamefont {M.~D.}\ \bibnamefont {Reid}},\
  }\href {\doibase 10.1103/PhysRevA.90.062337} {\bibfield  {journal} {\bibinfo
  {journal} {Phys. Rev. A}\ }\textbf {\bibinfo {volume} {90}},\ \bibinfo
  {pages} {062337} (\bibinfo {year} {2014})}\BibitemShut {NoStop}%
\bibitem [{\citenamefont {Li}\ \emph {et~al.}(2007)\citenamefont {Li},
  \citenamefont {Fei}, \citenamefont {Wang},\ and\ \citenamefont
  {Wu}}]{fei.pra.75.012311.2007}%
  \BibitemOpen
  \bibfield  {author} {\bibinfo {author} {\bibfnamefont {Z.-G.}\ \bibnamefont
  {Li}}, \bibinfo {author} {\bibfnamefont {S.-M.}\ \bibnamefont {Fei}},
  \bibinfo {author} {\bibfnamefont {Z.-X.}\ \bibnamefont {Wang}}, \ and\
  \bibinfo {author} {\bibfnamefont {K.}~\bibnamefont {Wu}},\ }\href {\doibase
  10.1103/PhysRevA.75.012311} {\bibfield  {journal} {\bibinfo  {journal} {Phys.
  Rev. A}\ }\textbf {\bibinfo {volume} {75}},\ \bibinfo {pages} {012311}
  (\bibinfo {year} {2007})}\BibitemShut {NoStop}%
\end{thebibliography}


%

\end{document}